\documentclass[%
reprint,
superscriptaddress,
showpacs,
preprintnumbers,
amsmath,amssymb,
aps,
prl,
]{revtex4-1}
\usepackage[colorinlistoftodos]{todonotes}
\usepackage{graphicx}
\usepackage{color}
\usepackage[caption=false]{subfig}
\usepackage{dcolumn}
\usepackage{bm}
\usepackage{hyperref}
\usepackage[export]{adjustbox}
\usepackage{subfig}
\usepackage[mathlines]{lineno}

\begin{document}

\preprint{APS/123-QED}

\title{Potential Flows with Electromagnetically-Induced Circulation in a Hele-Shaw Cell}

\author{Kyle I. McKee}
\email{kimckee@mit.edu}
\affiliation{Department of Mathematics, Massachusetts Institute of Technology, Cambridge, Massachusetts, USA}%


\author{John W.M. Bush}
\affiliation{Department of Mathematics, Massachusetts Institute of Technology, Cambridge, Massachusetts, USA}

\date{\today}

\begin{abstract}
In Hele-Shaw cells, pressure-driven viscous fluid motion between two closely-spaced plates 
gives rise to a two-dimensional potential flow with zero circulation.
Here, we show how the introduction of electromagnetic effects enables the realization of potential flows with circulation. We present canonical Hele-Shaw experiments with circulation prescribed by the electromagnetic configuration, and rationalize the observed flows theoretically. We also draw an analogy between this new class of circulatory potential flows and a class of electrostatic systems.
\end{abstract}

\maketitle

In 1898, \citet{hele1898flow} demonstrated experimentally that the motion of a viscous fluid through a narrow gap may be described in terms of a two-dimensional potential flow. \citet{stokes1898mathematical} later demonstrated the correspondence mathematically. Potential flow is usually associated with the inviscid, irrotational flows expected to arise at high Reynolds numbers. In practice, the potential flow description of high Reynolds number flow past obstacles fails since any small amount of viscosity induces vorticity in the boundary layer and concomitant flow separation \cite{prandtl1905uber,batchelor1967introduction}. Somewhat surprisingly, the most effective way to realize planar potential flows is in a thin gap geometry dominated by viscous effects \cite[pp. 8]{van1982album}, the so-called Hele-Shaw cell. 

In general, the two-dimensional potential flow past a set of obstacles is not unique until the circulation around each obstacle is prescribed. While the circulation alone dictates the lift on an aerofoil in an inviscid flow \citep[pp. 406]{batchelor2000}, the circulation cannot be predicted within the framework of inviscid theory, without appealing to additional physics. Ultimately, viscous effects dictate the lift in high Reynold's number flows. For steady flow past an aerofoil with a single sharp corner, the Kutta condition \citep[pp. 438]{batchelor2000} requires that the velocity remain finite at the trailing edge of an aerofoil and may be applied to specify a unique value of the circulation. In problems without sharp corners, there is no general procedure for computing the circulation \footnote{see \cite{gonzalez2022variational} for a recent attempt to generalize the Kutta condition}. For example, in the potential flow modeling of flow past a rotating cylinder, the Magnus lift \cite{magnus1853ueber} is given by $F_L=\rho U \Gamma$ \citep[pp. 427]{batchelor1967introduction}, and the circulation must be prescribed in an \textit{ad hoc} manner. Here, we demonstrate how one may introduce a predictable amount of circulation into a two-dimensional potential flow.

\begin{figure}
\centerline{\includegraphics[width=\linewidth]{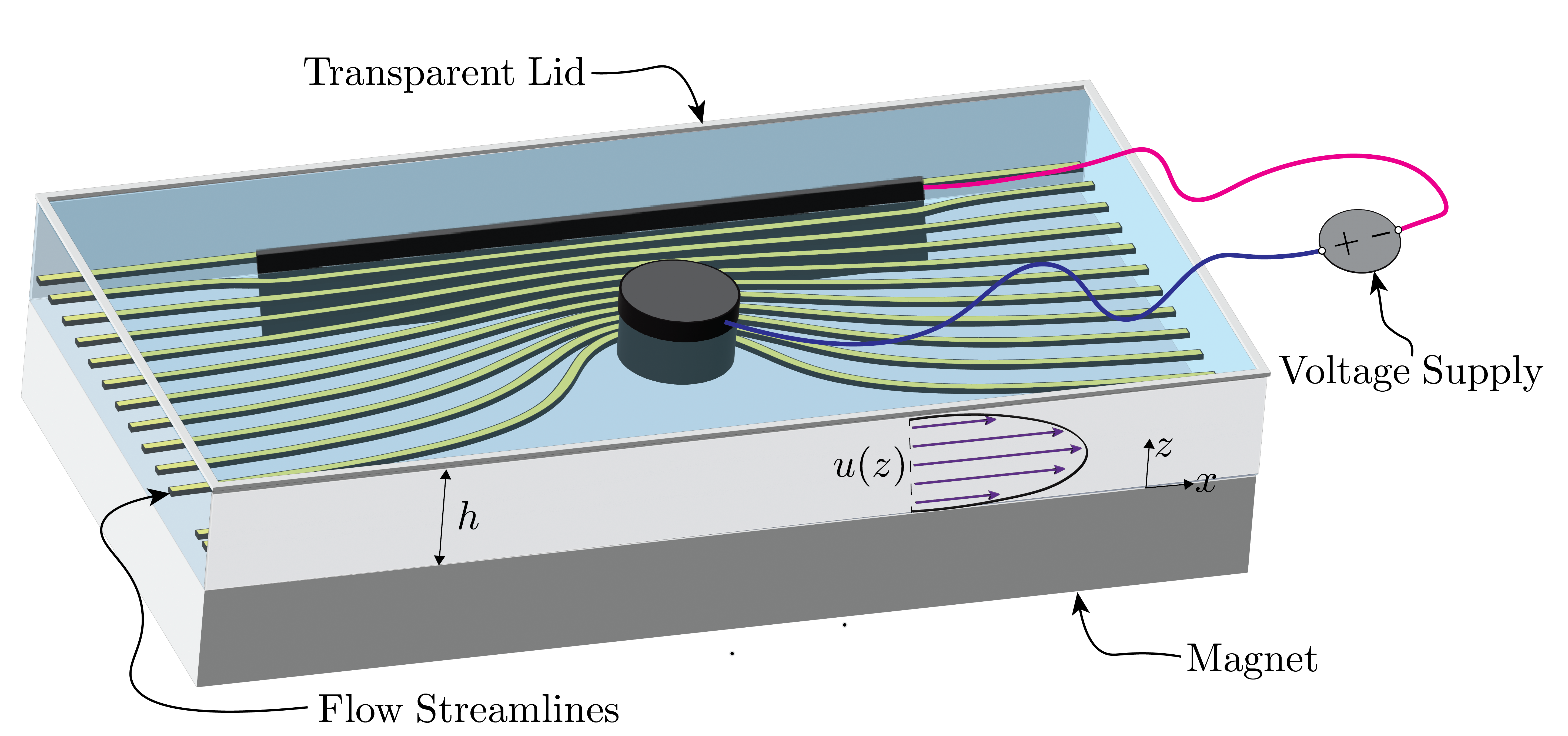}}
 \caption{\label{fig:schematic} The Hele-Shaw cell is filled with saltwater and placed atop a permanant neodymium magnet, whose field is $\boldsymbol{B}=B_0\hat{\boldsymbol{z}}$. Aluminum obstacles are placed into the cell from above. A pressure gradient drives flow from left to right. Circulation is induced by applying an electric current between the anode and cathode in the presence of the magnetic field.}
\end{figure}

In the Hele-Shaw cell, inertial terms are negligible and viscous forces are balanced by pressure gradients. Flow is fully developed across the gap height, specifically with a parabolic profile in the vertical $z$-coordinate, as depicted in figure \ref{fig:schematic}. In each plane of the cell, the two-dimensional fluid velocity $\boldsymbol{u}$ is proportional to the pressure gradient, $\nabla p$, so the incompressibility condition, $\nabla \cdot \boldsymbol{u}=0$, dictates that the pressure be harmonic, $\nabla^2p=0$. Thus, the fluid flow is a two-dimensional potential flow with the potential being the pressure. Notably, such fluid flows necessarily possess zero circulation, as follows from evaluating the circulation around any closed curve $\partial \Omega$, $\Gamma \propto \int_{\partial \Omega}\nabla p \cdot\boldsymbol{t}dl$. Since $p$ is a single-valued function, $\Gamma$ necessarily vanishes.
Pressure-driven Hele-Shaw flows are thus only capable of realizing potential flows with zero circulation. The potential flow solution realized in a pressure-driven Hele-Shaw cell is thus unambiguous: the circulation around any body in the flow is always precisely zero. 
Thus, the Hele-Shaw cell is limited in its capacity to realize potential flows. As stated by \citet[pp. 9]{van1982album}, ``\emph{The Hele-Shaw analogy cannot represent a flow with circulation. It therefore shows the streamlines of potential flow past an inclined plate with zero lift}''. 
We here demonstrate that electromagnetic effects may be used to realize Hele-Shaw flows with non-zero circulation.

The standard circulation-free Hele-Shaw cell 
serves 
as a robust platform for investigating problems in porous media and electrostatics. In the analogy with porous media flows obeying Darcy's Law \cite{darcy1856fontaines,brown2002henry,whitaker1986flow}, $\boldsymbol{u}=-k\nabla p$, the porous media permeability $k$ is mapped directly to $k=h^2/12$, $h$ being the Hele-Shaw cell thickness. Thus, Darcy flows with variable permeability can be modeled in Hele-Shaw cells by varying the gap thickness, $h(x)$. Porous media flows arise in filtration, petroleum production, and groundwater flows \cite{smith1969investigation}; consequently, the Hele-Shaw analogy is widely used in the laboratory modeling of flow in porous media  \cite{homsy1987viscous,singh2019capillary,juanes2020multiphase,qiu2023wetting}.
In the electrostatic analogy (see table \ref{tab:tableanalogies}), fluid flow streamlines correspond to electric field lines, impermeable obstacles to perfect electrical insulators, and fluid sources and sinks to positive and negative point charges, respectively. The electrostatic analogy is valuable in that it allows for the visual solution to electrostatic problems in complex geometries where it may be difficult to obtain mathematical solutions \citep[Fig. 9 and 10]{hele1898flow}. 

McKee \cite{mckee2024exact} recently showed theoretically that circulation may be introduced into Hele-Shaw flows, when electric current passes through an electrically conducting fluid in the presence of a uniform external magnetic field. A single experiment was presented to motivate the theoretical developments and  
no quantitative comparison with theory was presented. 
While the possibility of combining pressure-driven flows with electromagnetic flows was suggested in that paper, no such experiments were performed. We proceed by reporting a set of experiments in a Hele-Shaw cell that illustrate pressure driven potential flows with electomagnetically-induced circulation. We then compare our experimental observations to theoretical predictions.

A Hele-Shaw cell was constructed out of stacked laser-cut acrylic sheets, with a gap thickness of $h=1.5\mathrm{mm}$. The cell was filled with salt-water, 
with molarity $3.31\pm .16 \mathrm{mol/L}$, which at 20\textdegree C corresponds to a viscosity of $\mu=1.43\pm 0.07\mathrm{cP}$ \cite[pp. 5-142]{haynes2016crc}. The entire cell was placed atop a DZ08-N52 Neodymium Disc Magnet which produced a vertical magnetic field of $B_0=210\pm20\mathrm{mT}$, as measured with a Gaussmeter. The Hele-Shaw approximation requires that $h/L\ll 1$ and $\mathrm{Re}(h/L)=Uh\rho/\mu(h/L)\ll 1$, where $L$ is the characteristic lateral length-scale of the flow. In our case, using the typical obstacle scale $L=1\mathrm{cm}$, we find $h/L=0.15$ and $\mathrm{Re}(h/L)=0.09$. A schematic illustration of our experiment is presented in Fig. \ref{fig:schematic}.

\begin{figure}
 \centerline{\includegraphics[width=1\linewidth]{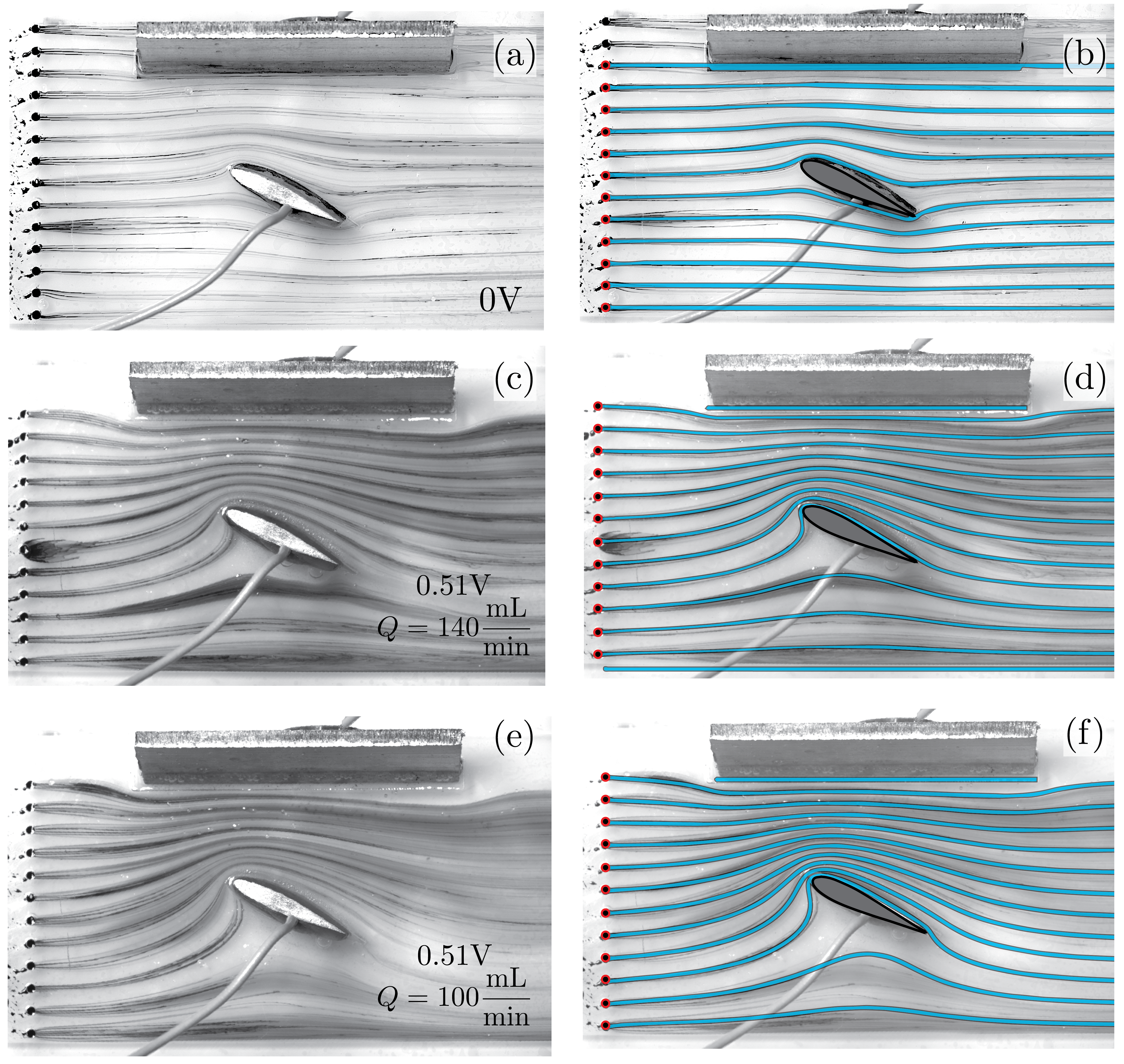}}
 \caption{\label{fig:aerofoil} Experimental images of Hele-Shaw flow past an aerofoil with: (a) zero circulation ($V_{\mathrm{app}}=0\mathrm{V}$), (c) circulation near the Kutta value ($Q=140\mathrm{mL/min}$ and $V_{\mathrm{app}}=0.51\mathrm{V}$), and (e) a circulation significantly larger than the Kutta value ($Q=100\mathrm{mL/min}$ and $V_{\mathrm{app}}=0.51\mathrm{V}$). The voltage was held constant and the flow rate adjusted so that $\Gamma/Q$ in (e) is 1.4 times larger than in (c). (b,d,f) Theoretical streamlines (blue) deduced from the analysis in Appendix A are overlaid on the experiments of (a,b,c). In (e), streamlines are fit to the experiments by choosing $\Gamma=1.22\Gamma_K$. In (f), theoretical streamlines corresponding to a 40\%  increase in the parameter $\Gamma/Q$ relative to panel (d).}
\end{figure}

\begin{figure}
\centerline{\includegraphics[width=1\linewidth]{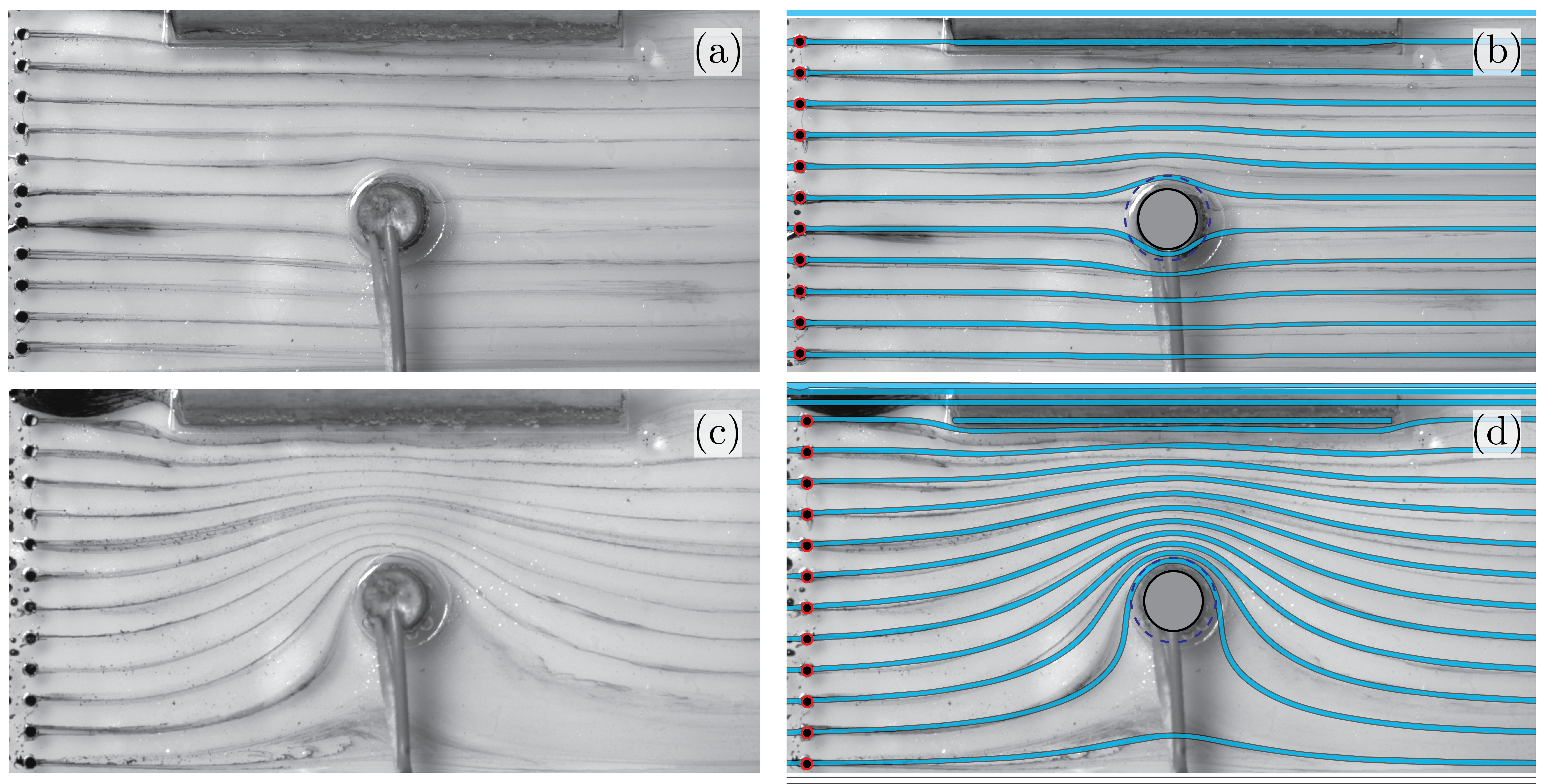} }
 \caption{\label{fig:circleexp} 
Experimental images of Hele-Shaw flow past an circle with: (a) zero circulation and (c) non-zero circulation induced by applying an electric current between the circle and the rectangular electrode. (b,d) Theoretical streamlines deduced from the analysis in Appendix A are overlaid on the experiments (a,c). In (d), the circulation was computed using (\ref{eq:circpred}).
The vertical magnetic field magnitude was measured to be $B_0=210\pm20\mathrm{mT}$, and the electrical current was $I=1.75\pm0.05\mathrm{mA}$.}
\end{figure}

The flow obstacles were made from aluminum and cut into the desired shapes using a CNC machine and then placed into the cell through a hole in the top plate of the Hele-Shaw cell \footnote{The hole in the top plate of the cell, through which the obstacles were placed, was made marginally larger than the obstacle dimensions to allow for bubbles to exit the cell in the event of electrolysis or chemical reactions taking place on the metal surface. Note that the voltage was never raised above 1 volt during the experiments to avoid electrolysis.}. The flows past an aerofoil and a circle are presented in figures \ref{fig:aerofoil} and \ref{fig:circleexp}, respectively. A thin aluminum rectangular electrode was also immersed in the flow to act as an electrical current sink, and is evident in the top of each panel of figures \ref{fig:aerofoil} and \ref{fig:circleexp}. A voltage was applied between the obstacles and the rectangular electrode, causing a net electrical current to flow between them, and inducing a flow circulation. Concurrently, a syringe pump was used to generate a uniform stream past the obstacles.

The first experiments were performed with an aerofoil-shaped obstacle. Figure \ref{fig:aerofoil}(a) illustrates a uniform stream flow past the aerofoil when no voltage was applied. In this case, no electromagnetic forces act on the fluid, resulting in a simple pressure-driven potential flow with zero circulation. As noted by Van Dyke \citep[Fig. 4, p. 10]{van1982album} ``\emph{because the Hele-Shaw flow cannot show circulation, the Kutta condition is not enforced at the trailing edge}''. 
We then relax the zero-circulation restriction using electromagnetic forcing. As the voltage between the aerofoil and the plate is increased, a non-zero circulation is induced in the potential flow.
The flow in Fig. \ref{fig:aerofoil}(b) was achieved by fixing the flow rate in the Hele-Shaw cell ($Q=140\mathrm{mL/s}$) and adjusting the voltage difference between the aerofoil and rectangular electrode until the streamlines 
approximated those of the Kutta solution. 
In Fig. \ref{fig:aerofoil}(e), the voltage and thus circulation was fixed at the same value as that in Fig. \ref{fig:aerofoil}(c), while the flow rate was decreased by 40\% to $Q=100\mathrm{mL/s}$, thereby enhancing the effect of the circulation. 

In a second set of experiments, we examined the flow past a circular obstacle and directly measured the electric current with an ammeter. Figure \ref{fig:circleexp}(a) shows a uniform flow past the circle with no electrical current or induced circulation. When the voltage difference between the circle and rectangular electrode was increased, the streamlines were bent as illustrated in panel \ref{fig:circleexp}(c). Over the course of the experiment, the electrical current was measured to be $I=1.75\pm0.05\mathrm{mA}$. We now describe the theory used to produce the theoretical streamline predictions presented in figures \ref{fig:aerofoil} and \ref{fig:circleexp}.

Consider the flow of an electrically neutral conducting fluid in the Hele-Shaw geometry, subject to an external constant magnetic field directed normal to the cell walls, $\boldsymbol{B}=B_0\hat{\boldsymbol{z}}$. Let the obstacles in the cell (aerofoils, \textit{etc.}) be electrodes to which voltages are applied (see Fig. \ref{fig:schematic}). Electric current then flows through the fluid according to Ohm's law, $\boldsymbol{J}=\sigma \boldsymbol{E}$, where $\boldsymbol{J}$ is the current density, $\boldsymbol{E}$ is the electric field, and $\sigma$ the electrical conductivity of the fluid. Then, the magnetic body force on the fluid is prescribed by the Lorentz force, $\boldsymbol{F}_{\mathrm{L}}=\sigma B_0 \boldsymbol{E} \times \hat{\boldsymbol{z}}$. Since $\nabla \boldsymbol{\cdot} \boldsymbol{F}_{\mathrm{L}}=\nabla \times \boldsymbol{F}_{\mathrm{L}}=0$, one may represent the magnetic force in terms of a magnetic potential $\phi$ as follows, $\boldsymbol{F}_{\mathrm{L}}=\nabla \phi(x,y)$ \cite{mckee2024exact,david2023magnetostokes}. As such, the total flow in the gap between $z=0$ and $z=h$ becomes,
\begin{equation}\label{eq:velz}
    \boldsymbol{u}(z,y,z)=\frac{z\left(h-z\right)}{2\mu}\left(\nabla \phi -\nabla p \right),
\end{equation}
where both the pressure, $p(z,y)$, and magnetic potential, $\phi(x,y)$, are harmonic functions \footnote{\citet{mckee2024exact} showed that the solution form (\ref{eq:velz}) is valid in the limit of small Hartmann numbers; when the Hartman number becomes too large, the parabolic profile in the vertical coordinate becomes spoiled owing to magnetic effects in the boundary layer \citep[figure 5.10 on pp. 153]{davidson2002introduction}}. Thus, in the presence of electromagnetic forcing, the flow retains a potential flow structure. Notably, while the pressure is constrained to be single-valued, there is no such restriction on the magnetic potential, $\phi$ \cite{mckee2024exact}, which introduces the possibility of non-zero circulation. We proceed by deriving the magnitude of the electromagnetically-induced circulation.

To simplify our discussion, we choose now to analyse the flow in the mid-plane of the Hele-Shaw cell where $z=h/2$ and we label this flow $\overline{\boldsymbol{u}}=\boldsymbol{u}(x,y,h/2)$. Note that the flow is identical in each horizontal plane, up to a scaling constant. In the mid-plane, the circulatory part of the flow can be written as
\begin{equation}
\overline{\boldsymbol{u}}_{\mathrm{circ}}=\frac{h^2}{8\mu}\nabla \phi=\frac{h^2B_0}{8\mu}\boldsymbol{J}\times \hat{\boldsymbol{z}},
\end{equation}
so that the induced circulation around a body $\Omega$ may be written
\begin{equation}
    \Gamma = \frac{h^2B_0}{8\mu}\int_{\partial \Omega}\boldsymbol{J}\times \hat{\boldsymbol{z}}\boldsymbol{\cdot}\hat{\boldsymbol{t}}dl,
\end{equation}
where $\hat{\boldsymbol{t}}$ is the unit tangent vector to $\partial \Omega$. 
Noting that the total current leaving the body is given by $I=h\int_{\partial \Omega}\boldsymbol{J}\boldsymbol{\cdot}\hat{\boldsymbol{n}}dl$, we may use the relation $\boldsymbol{J}\times \hat{\boldsymbol{z}}\boldsymbol{\cdot}\hat{\boldsymbol{t}}=-\boldsymbol{J}\boldsymbol{\cdot}\hat{\boldsymbol{n}}$ to write the induced flow circulation as
\begin{equation}\label{eq:circpred}
    \Gamma = -\frac{hB_0I}{8\mu},
\end{equation}
as measured in a right-handed sense about the positive $z-$axis \footnote{if $\hat{\boldsymbol{t}}=(t_x,t_y)$, then $\hat{\boldsymbol{n}}=(t_y,-t_x)$. It follows that $\boldsymbol{J}\times \hat{\boldsymbol{z}}\boldsymbol{\cdot}\hat{\boldsymbol{t}}=(J_yt_x-J_xt_y)$ is equivalent to $-\boldsymbol{J}\boldsymbol{\cdot}\hat{\boldsymbol{n}}=-J_xt_y+J_yt_x$.}. When there are multiple bodies in the flow, the circulation $\Gamma_k$ around the $k^\mathrm{th}$ body can be written in terms of the net current leaving that body, $I_k$, according to $\Gamma_k=-hB_0I_k/(8\mu)$.
The induced circulation around each electrode scales as the product of the magnetic field strength and the net current leaving that electrode. The magnitude of the circulation may thus be deduced in terms of a few quantities that are relatively straightforward to measure: $B_0$, $I$, $h$, and $\mu$. With the circulation around each obstacle prescribed by (\ref{eq:circpred}), the mathematical problem of finding the potential flow past the obstacles is fully posed. We now outline our technique for computing such flows, which makes use of a complex variables formulation of the problem \cite{mckee2024exact}.

\begin{table*}
\caption{\label{tab:tableanalogies}%
Summary of existing and new analogies for Hele-Shaw flows past perfect obstacles. Note that in addition to the boundary conditions displayed in the table, the circulations of the gradients of $p$, $V$ and $\psi$ must all be zero; specifically, $\int_{\partial \Omega_k} \nabla p\boldsymbol{\cdot}\boldsymbol{t}dl=\int_{\partial \Omega_k} \nabla V\boldsymbol{\cdot}\boldsymbol{t}dl=\int_{\partial \Omega_k} \nabla \psi \boldsymbol{\cdot}\boldsymbol{t}dl=0$ for any body $\partial \Omega_k$..
}
\begin{ruledtabular}
\begin{tabular}{lcccr}
\textrm{System\footnote{We use the abbreviations in the table: ``PD" denotes pressure-driven;``HSF" denotes Hele-Shaw flow;``ES" denotes electrostatic; "MS" denotes magnetohydrodynamic.}}&
\textrm{Field}&\textrm{Equation}&
\textrm{BC as $||\boldsymbol{x}||\rightarrow \infty$}&
\textrm{BC on $\partial \Omega_k$}\\
\colrule
PD HSF past obstacles&$\boldsymbol{u}=-\frac{h^2}{8\mu}\nabla p$&$\nabla^2 p=0$ & $\boldsymbol{u}\sim U\hat{\boldsymbol{x}}$ & $\boldsymbol{u}\boldsymbol{\cdot}\boldsymbol{n}=0$\\
ES field around insulators &$\boldsymbol{E}=-\nabla V $&$\nabla^2V=0$& $\boldsymbol{E}\sim E_0\hat{\boldsymbol{x}}$ & $\boldsymbol{E}\boldsymbol{\cdot}\boldsymbol{n}=0$\\
\hline
ES field around conductors &$\boldsymbol{E}=-\nabla V $&$\nabla^2 V=0$& $||\nabla V||\sim 0$ & $V=V_k$ \\
MH HSF around obstacles&$\boldsymbol{u}=\nabla \psi \times \hat{\boldsymbol{z}}$&$\nabla^2\psi=0$ & $||\nabla \psi||\sim 0$ & $\psi=h^2 B_0\sigma V_k/(8\mu)$ \\
\end{tabular}
\end{ruledtabular}
\end{table*}


The electrical potential $V$ ($\boldsymbol{E}=-\nabla V$) can be represented as the real part of a complex potential $W_E(z)$ that is analytic over the fluid domain, where $z=x+\mathrm{i}y$. The complex electric field is then given by $E=-\overline{dW_E/dz}$, where the overline represents complex conjugation and the current density is simply $J=\sigma E$. Within the complex variables framework, the cross product in the Lorentz force becomes a simple multiplication by $\mathrm{i}$. Meanwhile, the harmonic pressure field can also be represented as the real part of an analytic function $p=\mathrm{Re}\left\{W_{\mathrm{p}}\right\}$. The complex fluid flow velocity in the midplane becomes (see \citep{mckee2024exact} for details)
\begin{equation}\label{eq:comp_eq} u=\frac{h^2}{8\mu}\left(\overline{-\mathrm{i}\sigma B_0\frac{dW_E}{dz} - \frac{dW_{\mathrm{p}}}{dz} }\right).
\end{equation}
The complex potential for a source of circulation takes the form $\mathrm{i}\left(\Gamma/2\pi\right) \log{\left(z\right)}$ with $\Gamma\in\mathbb{R}$, whose real part is multi-valued. Hence, the physical constraint that $p$ be single-valued clearly precludes the existence of circulation. However, there is no such restriction on the magnetic potential as was evidenced in the calculation leading to (\ref{eq:circpred}). It has been shown that the complex potential can be found via conformal maps or rapidly converging series solutions \citep{mckee2024exact}. We use an extension of the framework laid out in \cite{mckee2024exact} to construct theoretical predictions for the experiments presented herein. Numerical details are given in Appendix A. We proceed by comparing theoretical predictions, as computed using our complex variables framework, with the experiments illustrated in figures \ref{fig:circleexp} and \ref{fig:aerofoil}.

Consider the aerofoil experiments presented in Fig. \ref{fig:aerofoil}. In Fig. \ref{fig:aerofoil}(a), no electrical current is applied and hence a zero circulation solution past the aerofoil is realized. Theoretical streamlines corresponding to $\Gamma=0$ are plotted in blue in Fig. \ref{fig:aerofoil}(b), yielding good agreement with the experiment. In the experiment presented in Fig. \ref{fig:aerofoil}(c), the flow circulation was tuned to roughly satisfy the Kutta condition. Details on how to compute theoretical flow streamlines corresponding to the precise Kutta condition are included in Appendix A. In the aerofoil experiments, the electrical current was not directly measured so that we cannot employ (\ref{eq:circpred}) to predict the circulation. Instead, we used $\Gamma$ to fit the streamlines and found an accurate theoretical fit for $\Gamma=1.22\Gamma_K$, 22\% above the Kutta condition. The theoretically predicted streamlines, 
corresponding to $\Gamma=1.22\Gamma_K$, are plotted in Fig. \ref{fig:aerofoil}(d), and show good agreement with the experiments. In the experiments of Fig. \ref{fig:aerofoil}(e), the circulation is kept the same as in the experiment in Fig. \ref{fig:aerofoil}(c), but the flow rate is reduced by a factor of 1.4 to $100 \mathrm{mL} / \mathrm{min}$. Since, for a particular geometry, the streamlines only depend on the non-dimensional parameter $G=\Gamma/(UL)$ with $L$ being the body length \footnote{$G$ is essentially a lift coefficient. Using the lift formula, it is seen that $G\propto F_L/(\frac{1}{2}\rho U^2 L)=2\Gamma/UL$}, this decrease in $U$ has the same affect on streamlines as an increase of $\Gamma$ by the same factor. 
Thus, the theoretical prediction for the streamlines at the new flow rate, in Fig. \ref{fig:aerofoil}(f), were obtained by  simply repeating the streamline calculation leading to Fig. \ref{fig:aerofoil}(d) but with a 40\% increase in the circulation,  $\Gamma=1.4(1.22\Gamma_K)=1.71\Gamma_K$, and without changing the flow rate. The theoretically predicted streamlines with $\Gamma = 1.71\Gamma_K$ are plotted in blue in panel \ref{fig:aerofoil}(f), and show excellent agreement with the experiment of panel \ref{fig:aerofoil}(e). To summarize, the circulation was fit to reproduce the experimental streamlines from Fig. \ref{fig:aerofoil}(c) and then used to predict the theoretical streamlines presented in Fig. \ref{fig:aerofoil}(f).

Now consider the experiments presented in Fig. \ref{fig:circleexp}, wherein the electric current was measured so that the validity of (\ref{eq:circpred}) could be assessed directly. In Fig. \ref{fig:circleexp}(a), no electrical current is applied and a circulation-free flow is produced. Theoretical streamlines corresponding to $\Gamma=0$ are plotted in Fig. \ref{fig:circleexp}(b), yielding good agreement with the experiments. In the experiment illustrated in Fig. \ref{fig:circleexp}(c), the electrical current between the circle and rectangular electrode was increased to induce flow circulation, and its value was measured with an ammeter. The circulation is then predicted using (\ref{eq:circpred}) and we find that $\Gamma=48\pm 7 \mathrm{mm^2/s}$, where the error is determined through summing the errors of $B_0,I,h$ and $\mu$ in quadrature. In Fig. \ref{fig:circleexp}(b), we draw the theoretically predicted streamlines corresponding to our computed circulation. Here, we chose to plot flow streamlines for a value of $\Gamma$ within the error range of our prediction, $\Gamma\in(41,55) \mathrm{mm^2/s}$, which best aligns with the experiments. The experimental and theoretical streamlines are in satisfactory agreement with a circulation value of $\Gamma=55\mathrm{mm^2/s}$.

We note that the Hele-Shaw approximation breaks down inside a boundary layer around each obstacle with a thickness on the order of $h$, since the potential flow solution fails to satisfy the no-slip condition on each obstacle surface \cite{klettner2022effect}. In Fig. \ref{fig:circleexp}(b) and \ref{fig:circleexp}(d), the dark blue dotted line outlines a boundary layer of thickness $h$ around the obstacle. The streamline that approaches the circle most closely in Fig. \ref{fig:circleexp}(a) enters this boundary layer wherein enhanced shear promotes diffusion of the dye, rendering the dye line invisible in the wake of the circle. 
Scaling up the obstacles size and increasing the dye line separation from obstacles would mitigate this effect; however, our experiment was limited by the size of our magnet, which was roughly $8\mathrm{cm}$ in diameter. 



We have shown how electromagnetic effects allow one to introduce circulation into Hele-Shaw flows; furthermore, we have demonstrated that the magnitude of this circulation is simply described by (\ref{eq:circpred}). We have illustrated this conceptual advance with Hele-Shaw flows past aerofoils and circles, which conformed to our theoretical predictions. Finally, we show how these new Hele-Shaw flows, which possess circulation, allow for an extension of the existing analogy with electrostatics. The existing and new analogies are summarized in table \ref{tab:tableanalogies}. In the traditional pressure-driven Hele-Shaw cell, flow streamlines around obstacles correspond to electric field lines impinging upon perfect electrical \emph{insulators}. In the magnetohydrodynamic Hele-shaw cell developed here, streamlines correspond to contours of electrical potential around a set of perfect electrical \emph{conductors}.
When the pressure vanishes ($W_p=0$), it is evident from (\ref{eq:comp_eq}) that the complex velocity potential is given by  $W_U=\phi+\mathrm{i}\psi=\mathrm{i}h^2\sigma B_0/(8\mu) W_E$, where $\phi$ is the velocity potential and $\psi$ the streamfunction. The condition $W_p=0$ is satisfied whenever all obstacles in the flow are perfect electrical conductors and when no external pressure gradient is applied. Hence, the streamfunction of the fluid flow satisfies $\psi=h^2 B_0\sigma V/(8\mu)$, where $V$ is the electrical potential. Dye lines in the magnetohydrodynamic Hele-Shaw cell thus trace out the electrical potential contours for the electrostatic solution around conducting obstacles with applied voltages.

\section{Acknowledgements}
\vspace{5mm}
K.M. thanks Simon Rufer and Jon Bessette for help with the fabrication of the aluminum electrodes. K.M. also thanks Darren Crowdy, Yu Qiu, and Bauyrzhan Primkulov for useful discussions. 

\bibliography{references7.bib}

\begin{thebibliography}{29}%
\makeatletter
\providecommand \@ifxundefined [1]{%
 \@ifx{#1\undefined}
}%
\providecommand \@ifnum [1]{%
 \ifnum #1\expandafter \@firstoftwo
 \else \expandafter \@secondoftwo
 \fi
}%
\providecommand \@ifx [1]{%
 \ifx #1\expandafter \@firstoftwo
 \else \expandafter \@secondoftwo
 \fi
}%
\providecommand \natexlab [1]{#1}%
\providecommand \enquote  [1]{``#1''}%
\providecommand \bibnamefont  [1]{#1}%
\providecommand \bibfnamefont [1]{#1}%
\providecommand \citenamefont [1]{#1}%
\providecommand \href@noop [0]{\@secondoftwo}%
\providecommand \href [0]{\begingroup \@sanitize@url \@href}%
\providecommand \@href[1]{\@@startlink{#1}\@@href}%
\providecommand \@@href[1]{\endgroup#1\@@endlink}%
\providecommand \@sanitize@url [0]{\catcode `\\12\catcode `\$12\catcode `\&12\catcode `\#12\catcode `\^12\catcode `\_12\catcode `\%12\relax}%
\providecommand \@@startlink[1]{}%
\providecommand \@@endlink[0]{}%
\providecommand \url  [0]{\begingroup\@sanitize@url \@url }%
\providecommand \@url [1]{\endgroup\@href {#1}{\urlprefix }}%
\providecommand \urlprefix  [0]{URL }%
\providecommand \Eprint [0]{\href }%
\providecommand \doibase [0]{http://dx.doi.org/}%
\providecommand \selectlanguage [0]{\@gobble}%
\providecommand \bibinfo  [0]{\@secondoftwo}%
\providecommand \bibfield  [0]{\@secondoftwo}%
\providecommand \translation [1]{[#1]}%
\providecommand \BibitemOpen [0]{}%
\providecommand \bibitemStop [0]{}%
\providecommand \bibitemNoStop [0]{.\EOS\space}%
\providecommand \EOS [0]{\spacefactor3000\relax}%
\providecommand \BibitemShut  [1]{\csname bibitem#1\endcsname}%
\let\auto@bib@innerbib\@empty
\bibitem [{\citenamefont {Hele-Shaw}(1898)}]{hele1898flow}%
  \BibitemOpen
  \bibfield  {author} {\bibinfo {author} {\bibfnamefont {H.~S.}\ \bibnamefont {Hele-Shaw}},\ }\href@noop {} {\bibfield  {journal} {\bibinfo  {journal} {Nature}\ }\textbf {\bibinfo {volume} {58}},\ \bibinfo {pages} {34} (\bibinfo {year} {1898})}\BibitemShut {NoStop}%
\bibitem [{\citenamefont {Stokes}(1898)}]{stokes1898mathematical}%
  \BibitemOpen
  \bibfield  {author} {\bibinfo {author} {\bibfnamefont {G.~G.}\ \bibnamefont {Stokes}},\ }\href@noop {} {\bibfield  {journal} {\bibinfo  {journal} {Brit. Ass. Rep}\ }\textbf {\bibinfo {volume} {143}} (\bibinfo {year} {1898})}\BibitemShut {NoStop}%
\bibitem [{\citenamefont {Prandtl}(1905)}]{prandtl1905uber}%
  \BibitemOpen
  \bibfield  {author} {\bibinfo {author} {\bibfnamefont {L.}~\bibnamefont {Prandtl}},\ }\href@noop {} {\bibfield  {journal} {\bibinfo  {journal} {Verhandl. 3rd Int. Math. Kongr. Heidelberg (1904), Leipzig}\ } (\bibinfo {year} {1905})}\BibitemShut {NoStop}%
\bibitem [{\citenamefont {Batchelor}(1967)}]{batchelor1967introduction}%
  \BibitemOpen
  \bibfield  {author} {\bibinfo {author} {\bibfnamefont {G.~K.}\ \bibnamefont {Batchelor}},\ }\href@noop {} {\emph {\bibinfo {title} {An introduction to fluid dynamics}}}\ (\bibinfo  {publisher} {Cambridge University Press},\ \bibinfo {year} {1967})\BibitemShut {NoStop}%
\bibitem [{\citenamefont {Van~Dyke}(1982)}]{van1982album}%
  \BibitemOpen
  \bibfield  {author} {\bibinfo {author} {\bibfnamefont {M.}~\bibnamefont {Van~Dyke}},\ }\href@noop {} {\emph {\bibinfo {title} {An album of fluid motion}}}\ (\bibinfo  {publisher} {Parabolic Press Stanford},\ \bibinfo {year} {1982})\BibitemShut {NoStop}%
\bibitem [{\citenamefont {Batchelor}(2000)}]{batchelor2000}%
  \BibitemOpen
  \bibfield  {author} {\bibinfo {author} {\bibfnamefont {G.}~\bibnamefont {Batchelor}},\ }\href@noop {} {\emph {\bibinfo {title} {An introduction to fluid dynamics}}}\ (\bibinfo  {publisher} {Cambridge university press},\ \bibinfo {year} {2000})\BibitemShut {NoStop}%
\bibitem [{Note1()}]{Note1}%
  \BibitemOpen
  \bibinfo {note} {See \cite {gonzalez2022variational} for a recent attempt to generalize the Kutta condition}\BibitemShut {NoStop}%
\bibitem [{\citenamefont {Magnus}(1853)}]{magnus1853ueber}%
  \BibitemOpen
  \bibfield  {author} {\bibinfo {author} {\bibfnamefont {G.}~\bibnamefont {Magnus}},\ }\href@noop {} {\bibfield  {journal} {\bibinfo  {journal} {Annalen der physik}\ }\textbf {\bibinfo {volume} {164}},\ \bibinfo {pages} {1} (\bibinfo {year} {1853})}\BibitemShut {NoStop}%
\bibitem [{\citenamefont {Darcy}(1856)}]{darcy1856fontaines}%
  \BibitemOpen
  \bibfield  {author} {\bibinfo {author} {\bibfnamefont {H.}~\bibnamefont {Darcy}},\ }\href@noop {} {\emph {\bibinfo {title} {Les fontaines publiques de la ville de Dijon: exposition et application des principes {\`a} suivre et des formules {\`a} employer dans les questions de distribution d'eau}}},\ Vol.~\bibinfo {volume} {1}\ (\bibinfo  {publisher} {Victor dalmont},\ \bibinfo {year} {1856})\BibitemShut {NoStop}%
\bibitem [{\citenamefont {Brown}(2002)}]{brown2002henry}%
  \BibitemOpen
  \bibfield  {author} {\bibinfo {author} {\bibfnamefont {G.}~\bibnamefont {Brown}},\ }\href@noop {} {\bibfield  {journal} {\bibinfo  {journal} {Water Resources Research}\ }\textbf {\bibinfo {volume} {38}},\ \bibinfo {pages} {11} (\bibinfo {year} {2002})}\BibitemShut {NoStop}%
\bibitem [{\citenamefont {Whitaker}(1986)}]{whitaker1986flow}%
  \BibitemOpen
  \bibfield  {author} {\bibinfo {author} {\bibfnamefont {S.}~\bibnamefont {Whitaker}},\ }\href@noop {} {\bibfield  {journal} {\bibinfo  {journal} {Transport in porous media}\ }\textbf {\bibinfo {volume} {1}},\ \bibinfo {pages} {3} (\bibinfo {year} {1986})}\BibitemShut {NoStop}%
\bibitem [{\citenamefont {Smith}\ and\ \citenamefont {Greenkorn}(1969)}]{smith1969investigation}%
  \BibitemOpen
  \bibfield  {author} {\bibinfo {author} {\bibfnamefont {R.}~\bibnamefont {Smith}}\ and\ \bibinfo {author} {\bibfnamefont {R.}~\bibnamefont {Greenkorn}},\ }\href@noop {} {\bibfield  {journal} {\bibinfo  {journal} {Society of Petroleum Engineers Journal}\ }\textbf {\bibinfo {volume} {9}},\ \bibinfo {pages} {434} (\bibinfo {year} {1969})}\BibitemShut {NoStop}%
\bibitem [{\citenamefont {Homsy}(1987)}]{homsy1987viscous}%
  \BibitemOpen
  \bibfield  {author} {\bibinfo {author} {\bibfnamefont {G.~M.}\ \bibnamefont {Homsy}},\ }\href@noop {} {\bibfield  {journal} {\bibinfo  {journal} {Annual review of fluid mechanics}\ }\textbf {\bibinfo {volume} {19}},\ \bibinfo {pages} {271} (\bibinfo {year} {1987})}\BibitemShut {NoStop}%
\bibitem [{\citenamefont {Singh}\ \emph {et~al.}(2019)\citenamefont {Singh}, \citenamefont {Jung}, \citenamefont {Brinkmann},\ and\ \citenamefont {Seemann}}]{singh2019capillary}%
  \BibitemOpen
  \bibfield  {author} {\bibinfo {author} {\bibfnamefont {K.}~\bibnamefont {Singh}}, \bibinfo {author} {\bibfnamefont {M.}~\bibnamefont {Jung}}, \bibinfo {author} {\bibfnamefont {M.}~\bibnamefont {Brinkmann}}, \ and\ \bibinfo {author} {\bibfnamefont {R.}~\bibnamefont {Seemann}},\ }\href@noop {} {\bibfield  {journal} {\bibinfo  {journal} {Annual Review of Fluid Mechanics}\ }\textbf {\bibinfo {volume} {51}},\ \bibinfo {pages} {429} (\bibinfo {year} {2019})}\BibitemShut {NoStop}%
\bibitem [{\citenamefont {Juanes}\ \emph {et~al.}(2020)\citenamefont {Juanes}, \citenamefont {Meng},\ and\ \citenamefont {Primkulov}}]{juanes2020multiphase}%
  \BibitemOpen
  \bibfield  {author} {\bibinfo {author} {\bibfnamefont {R.}~\bibnamefont {Juanes}}, \bibinfo {author} {\bibfnamefont {Y.}~\bibnamefont {Meng}}, \ and\ \bibinfo {author} {\bibfnamefont {B.~K.}\ \bibnamefont {Primkulov}},\ }\href@noop {} {\bibfield  {journal} {\bibinfo  {journal} {Physical Review Fluids}\ }\textbf {\bibinfo {volume} {5}},\ \bibinfo {pages} {110516} (\bibinfo {year} {2020})}\BibitemShut {NoStop}%
\bibitem [{\citenamefont {Qiu}\ \emph {et~al.}(2023)\citenamefont {Qiu}, \citenamefont {Xu}, \citenamefont {Pahlavan},\ and\ \citenamefont {Juanes}}]{qiu2023wetting}%
  \BibitemOpen
  \bibfield  {author} {\bibinfo {author} {\bibfnamefont {Y.}~\bibnamefont {Qiu}}, \bibinfo {author} {\bibfnamefont {K.}~\bibnamefont {Xu}}, \bibinfo {author} {\bibfnamefont {A.~A.}\ \bibnamefont {Pahlavan}}, \ and\ \bibinfo {author} {\bibfnamefont {R.}~\bibnamefont {Juanes}},\ }\href@noop {} {\bibfield  {journal} {\bibinfo  {journal} {Proceedings of the National Academy of Sciences}\ }\textbf {\bibinfo {volume} {120}},\ \bibinfo {pages} {e2303515120} (\bibinfo {year} {2023})}\BibitemShut {NoStop}%
\bibitem [{\citenamefont {McKee}(2024)}]{mckee2024exact}%
  \BibitemOpen
  \bibfield  {author} {\bibinfo {author} {\bibfnamefont {K.~I.}\ \bibnamefont {McKee}},\ }\href {\doibase 10.1017/jfm.2024.618} {\bibfield  {journal} {\bibinfo  {journal} {Journal of Fluid Mechanics}\ }\textbf {\bibinfo {volume} {993}},\ \bibinfo {pages} {A11} (\bibinfo {year} {2024})}\BibitemShut {NoStop}%
\bibitem [{\citenamefont {Haynes}(2015)}]{haynes2016crc}%
  \BibitemOpen
  \bibfield  {author} {\bibinfo {author} {\bibfnamefont {W.~M.}\ \bibnamefont {Haynes}},\ }\href@noop {} {\emph {\bibinfo {title} {CRC handbook of chemistry and physics}}},\ \bibinfo {edition} {95th}\ ed.\ (\bibinfo  {publisher} {CRC press},\ \bibinfo {year} {2015})\BibitemShut {NoStop}%
\bibitem [{Note2()}]{Note2}%
  \BibitemOpen
  \bibinfo {note} {The hole in the top plate of the cell, through which the obstacles were placed, was made marginally larger than the obstacle dimensions to allow for bubbles to exit the cell in the event of electrolysis or chemical reactions taking place on the metal surface. Note that the voltage was never raised above 1 volt during the experiments to avoid electrolysis.}\BibitemShut {Stop}%
\bibitem [{\citenamefont {David}\ \emph {et~al.}(2024)\citenamefont {David}, \citenamefont {Hester}, \citenamefont {Xu},\ and\ \citenamefont {Aurnou}}]{david2023magnetostokes}%
  \BibitemOpen
  \bibfield  {author} {\bibinfo {author} {\bibfnamefont {C.~S.}\ \bibnamefont {David}}, \bibinfo {author} {\bibfnamefont {E.~W.}\ \bibnamefont {Hester}}, \bibinfo {author} {\bibfnamefont {Y.}~\bibnamefont {Xu}}, \ and\ \bibinfo {author} {\bibfnamefont {J.~M.}\ \bibnamefont {Aurnou}},\ }\href@noop {} {\bibfield  {journal} {\bibinfo  {journal} {Journal of Fluid Mechanics}\ }\textbf {\bibinfo {volume} {996}},\ \bibinfo {pages} {A33} (\bibinfo {year} {2024})}\BibitemShut {NoStop}%
\bibitem [{Note3()}]{Note3}%
  \BibitemOpen
  \bibinfo {note} {\protect \citet {mckee2024exact} showed that the solution form (\ref {eq:velz}) is valid in the limit of small Hartmann numbers; when the Hartman number becomes too large, the parabolic profile in the vertical coordinate becomes spoiled owing to magnetic effects in the boundary layer \protect \citep [figure 5.10 on pp. 153]{davidson2002introduction}}\BibitemShut {NoStop}%
\bibitem [{Note4()}]{Note4}%
  \BibitemOpen
  \bibinfo {note} {If $\protect \hat {\protect \bm {t}}=(t_x,t_y)$, then $\protect \hat {\protect \bm {n}}=(t_y,-t_x)$. It follows that $\protect \bm {J}\times \protect \hat {\protect \bm {z}}\protect \bm {\cdot }\protect \hat {\protect \bm {t}}=(J_yt_x-J_xt_y)$ is equivalent to $-\protect \bm {J}\protect \bm {\cdot }\protect \hat {\protect \bm {n}}=-J_xt_y+J_yt_x$.}\BibitemShut {Stop}%
\bibitem [{Note5()}]{Note5}%
  \BibitemOpen
  \bibinfo {note} {$G$ is essentially a lift coefficient. Using the lift formula, it is seen that $G\propto F_L/(\protect \frac {1}{2}\rho U^2 L)=2\Gamma /UL$}\BibitemShut {NoStop}%
\bibitem [{\citenamefont {Klettner}\ and\ \citenamefont {Smith}(2022)}]{klettner2022effect}%
  \BibitemOpen
  \bibfield  {author} {\bibinfo {author} {\bibfnamefont {C.}~\bibnamefont {Klettner}}\ and\ \bibinfo {author} {\bibfnamefont {F.}~\bibnamefont {Smith}},\ }\href@noop {} {\bibfield  {journal} {\bibinfo  {journal} {Journal of Fluid Mechanics}\ }\textbf {\bibinfo {volume} {934}},\ \bibinfo {pages} {A8} (\bibinfo {year} {2022})}\BibitemShut {NoStop}%
\bibitem [{\citenamefont {Gonzalez}\ and\ \citenamefont {Taha}(2022)}]{gonzalez2022variational}%
  \BibitemOpen
  \bibfield  {author} {\bibinfo {author} {\bibfnamefont {C.}~\bibnamefont {Gonzalez}}\ and\ \bibinfo {author} {\bibfnamefont {H.}~\bibnamefont {Taha}},\ }\href@noop {} {\bibfield  {journal} {\bibinfo  {journal} {J. Fluid Mech.}\ }\textbf {\bibinfo {volume} {941}} (\bibinfo {year} {2022})}\BibitemShut {NoStop}%
\bibitem [{\citenamefont {Davidson}(2001)}]{davidson2002introduction}%
  \BibitemOpen
  \bibfield  {author} {\bibinfo {author} {\bibfnamefont {P.~A.}\ \bibnamefont {Davidson}},\ }\href@noop {} {\emph {\bibinfo {title} {An introduction to magnetohydrodynamics}}}\ (\bibinfo  {publisher} {Cambridge University Press},\ \bibinfo {year} {2001})\BibitemShut {NoStop}%
\bibitem [{\citenamefont {Baddoo}\ \emph {et~al.}(2020)\citenamefont {Baddoo}, \citenamefont {Kurt}, \citenamefont {Ayton},\ and\ \citenamefont {Moored}}]{baddoo2020exact}%
  \BibitemOpen
  \bibfield  {author} {\bibinfo {author} {\bibfnamefont {P.~J.}\ \bibnamefont {Baddoo}}, \bibinfo {author} {\bibfnamefont {M.}~\bibnamefont {Kurt}}, \bibinfo {author} {\bibfnamefont {L.~J.}\ \bibnamefont {Ayton}}, \ and\ \bibinfo {author} {\bibfnamefont {K.~W.}\ \bibnamefont {Moored}},\ }\href@noop {} {\bibfield  {journal} {\bibinfo  {journal} {Journal of Fluid Mechanics}\ }\textbf {\bibinfo {volume} {891}},\ \bibinfo {pages} {R2} (\bibinfo {year} {2020})}\BibitemShut {NoStop}%
\bibitem [{\citenamefont {Crowdy}(2020)}]{crowdy2020solving}%
  \BibitemOpen
  \bibfield  {author} {\bibinfo {author} {\bibfnamefont {D.}~\bibnamefont {Crowdy}},\ }\href@noop {} {\emph {\bibinfo {title} {Solving problems in multiply connected domains}}}\ (\bibinfo  {publisher} {SIAM},\ \bibinfo {year} {2020})\BibitemShut {NoStop}%
\bibitem [{\citenamefont {Trefethen}(2018)}]{trefethen2018series}%
  \BibitemOpen
  \bibfield  {author} {\bibinfo {author} {\bibfnamefont {L.~N.}\ \bibnamefont {Trefethen}},\ }\href@noop {} {\bibfield  {journal} {\bibinfo  {journal} {The ANZIAM Journal}\ }\textbf {\bibinfo {volume} {60}},\ \bibinfo {pages} {1} (\bibinfo {year} {2018})}\BibitemShut {NoStop}%
\end{thebibliography}%
\newpage 
\section{Appendix A: Numerical Solution for Potential Flow and the Kutta Condition Imposition}
\subsection{The Conformal Map to Our Aerofoil}
\subsubsection{The Mapping}
\begin{figure}
\centerline{\includegraphics[width=0.77\linewidth]{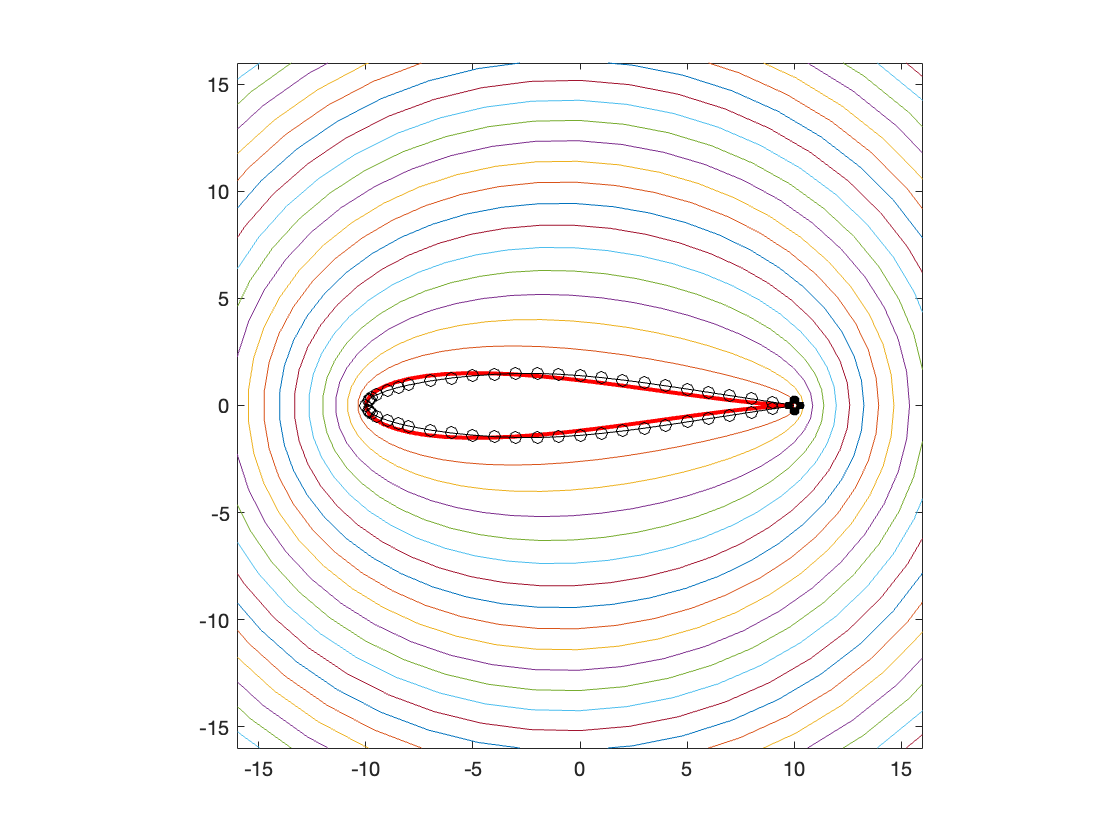}}
 \caption{\label{fig:aeromap}The aluminum aerofoil was cut using a CNC machine according to the points labelled by black circles connected by a thin black line. The fit of the experimental aerofoil to the analytical aerofoil expression described in equations (\ref{eq:zmap})-(\ref{eq:muval}) is given in red. The axes are labelled in units of millimeters. Despite its apparently complex form, the analytical map is parametrized according to the single parameter $\mu$, which we take to be equal to 0.78.}
\end{figure}

Our aluminum aerofoil is well approximated by the conformal map, from the unit circle, defined by
\begin{eqnarray}\label{eq:zmap}
z(\zeta ) & = & Se^{\mathrm{i}\theta}f(\zeta)+C,\\
f(\zeta)&=&\left(\zeta+s\right)+\frac{\mu}{\left(\zeta+s\right)},\\
s & = & \sqrt{\mu}-1,\\
S&=&20/\left(f(1)-f(-1)\right)\\
C&=&10-Sf(\zeta)\\
\mu & = & 0.78\label{eq:muval},
\end{eqnarray}
where the scale $S$ is chosen to give the aerofoil a length of $20\mathrm{mm}$. The choice of $s$ ensures the aerofoil has a sharp trailing edge at the image of the point $\zeta=1$. With $S$ and $s$ so chosen, based on the length and sharp edge of the aerofoil used in our experiment, the entire map is controlled by the single parameter $\mu \in \mathbb{R}$. We find $\mu \approx 0.78$ best represents our aerofoil (see Fig. \ref{fig:aeromap}). In our experiments, we have a pitch angle of $\theta=-25^{\circ}$.
\subsubsection{Flow Past the Isolated Aerofoil}
The flow exterior to the aerofoil can be found by first deducing a solution exterior to the unit disk and then using the conformal mapping principle. A uniform flow past a circle with free-stream velocity $U\in\mathbb{R}$ at an angle $\alpha$, and a circulation $\Gamma$, can be written as
\begin{equation}    W(\zeta)=U\left(e^{\mathrm{i}\alpha}\zeta+\frac{e^{-\mathrm{i}\alpha}}{\zeta}\right)+\frac{\mathrm{i}\Gamma}{2\pi}\log{\left(\zeta\right)}
\end{equation}
To ensure that the potential in the physical domain, $W(\zeta(z))$, has magnitude $U_0$ parallel to the real axis, we require $W \sim U_0 z$ as $|z|\rightarrow \infty$. We must thus choose $\alpha=\theta$ and $U=U_0 S$. The complex potential then becomes
\begin{equation}    W(\zeta)=U_0 S\left(\left(e^{\mathrm{i}\theta}\zeta+\frac{e^{-\mathrm{i}\theta}}{\zeta}\right)+\frac{\mathrm{i}\Gamma}{2\pi U_0 S}\log{\left(\zeta\right)}\right).
\end{equation}
Since the conformal map $z(\zeta)$ is independent of $\Gamma$ and $U_0$, it is clear that the streamlines of the flow are governed by the single non-dimensional parameter $\Gamma/\left(U_0 S\right)$ where $S$ is a constant for our single aerofoil geometry. Thus the ratio of circulation to free-stream velocity alone dictates sets the streamline shape around the aerofoil.
\subsubsection{The Kutta Condition Without Boundaries}
The velocity in the physical domain is given by $\left(\mathrm{d}W/\mathrm{d}\zeta\right)/\left(\mathrm{d}z/\mathrm{d}\zeta\right)$. If the map derivative $\mathrm{d}z/d\zeta$ possesses a zero, then the velocity diverges unless $\mathrm{d}W/\mathrm{d}\zeta$ possesses a zero simultaneously. The Kutta condition thus prescribes the circulation $\Gamma$ to force $\mathrm{d}W/\mathrm{d}\zeta$ to have a zero at exactly the same point as the zero of $\mathrm{d}z/d\zeta$ and so desingularizes the velocity. 

The map derivative for our aerofoil possesses a single zero at the point $\zeta=1$. To force the complex potential to have a zero derivative at the same point, we must solve
\begin{equation}
    0=\frac{\mathrm{d}W}{\mathrm{d}\zeta}\Big|_{\zeta=1}=U_0 S\left(\left(e^{\mathrm{i}\theta}-e^{-\mathrm{i}\theta}\right)+\frac{\mathrm{i}\Gamma}{2\pi U_0 S}\right),
\end{equation}
which yields the expression for the Kutta circulation
\begin{equation}\label{eq:kutt}
    \Gamma_K=-4\pi U_0 S\sin{\left(\theta\right)}.
\end{equation}

If there exist other obstacles in the flow, the Kutta condition changes and it cannot in general be computed analytically. In recent work, the Kutta condition near a flat wall was investigated analytically in the context of the ground effect \cite{baddoo2020exact}, through the use of the prime function machinery laid out by \citet{crowdy2020solving}.

\subsection{Numerical Method Accounting for Boundaries}
Our system does not consist of an isolated aerofoil in a uniform stream. The Hele-Shaw cell walls along with the rectangular electrode represent impermeable boundaries to the fluid flow. Compared to the isolated aerofoil, the effect of boundaries is to 1) alter the flow streamlines and 2) change the value of the Kutta circulation. The latter effect is well known in the context of the so-called ground effect, whereby an aerofoil near a flat wall experiences an enhanced lift. The Kutta circulation in the ground effect was studied analytically by \citet{baddoo2020exact} for the special class of doubly-connected geometries which does not include our system. We are thus required to solve numerically for the flow streamlines in our system.

In order to do so, we exploit the complex variables formulation of potential flow, wherein the complex velocity is given by $u=\overline{dW/dz}$ for a complex analytic function $W(z)$. We denote each obstacle (the aerofoil, the rectangular electrode, and the upper and lower walls) by $\Omega_k$ where $k\in\{1,2,3,4\}$, and the corresponding boundaries by $\partial \Omega_k$. In our scheme, the upper and lower walls are represented by horizontal slits much longer than the aerofoil length. At infinity, the complex potential must converge to a uniform stream so that,
\begin{equation}\label{eq:Wbvp}
    W \sim \overline{U}z,\;|z|\rightarrow \infty,
\end{equation}
and along each body boundary, the impermeability condition reduces to
\begin{equation}\label{eq:Wbvp2}
    \mathrm{Im}\left\{W(z)\right\}=\psi_k,\;z\in \partial \Omega_k.
\end{equation}
The Laplace problem is not completely posed until the circulation is also specified around each boundary,
\begin{eqnarray}\label{eq:Wbvp3}
    \int_{\partial \Omega_k}\mathrm{Re}\left\{\frac{dW}{dz}\frac{dz}{ds}\right\}ds=\Delta \mathrm{Re}\{W\}=\Gamma_k,
\end{eqnarray}
for each $k\in\{1,2,3,4\}$. When bodies are smooth obstacles without sharp corners or thin appendages, the solution may be expressed in terms sum of Laurent series centered inside each of the obstacles \cite{trefethen2018series}. However, when obstacles are arc segments or if they possess sharp edges (such as the aerofoil considered herein), the shape needs to be expanded via a conformal map first and then the Laurent series centered in the new domain. In our problem, the conformal maps $\zeta_j(z)$ between each object exterior and the exterior of the unit circle is known. For each slit, this map is the usual Joukowsky map whereas the aerofoil map is given by (\ref{eq:zmap}). Thus, we can represent the complex potential in a series,
\begin{equation}\label{eq:Wform}
    W(z)  =  \overline{U}z+a_0 + \sum_{j=1}^{N}\left(\sum_{k=1}^{\infty} \frac{a_k^{(j)}}{\zeta_j^k} + \mathrm{i}\frac{\Gamma_j}{2\pi} \log{\left(\zeta_j\right)}\right).
\end{equation}
We note that in our physical context, we know that circulation around the aerofoil ($j=1$) must be equal and opposite to that around the rectangular electrode ($j=2$) since all the electric current exiting the aerofoil enters the electrode. Thus, we can write the truncated complex potential in terms of a single circulation, $\Gamma$, as follows
\begin{equation}\label{eq:Wform1}
    W(z)  \approx  \overline{U}z+a_0 + \sum_{j=1}^{4}\left(\sum_{k=1}^{N_L} \frac{a_k^{(j)}}{\zeta_j^k} + \mathrm{i}\frac{\Gamma}{2\pi} \log{\left(\zeta_1/\zeta_2\right)}\right).
\end{equation}
It is straightforward to truncate each Laurent series and obtain the coefficients  $a_k^{(j)}$ if $\Gamma$ is known a priori. After choosing sample points along each boundary, enforcing the impermeability condition (\ref{eq:Wbvp2}) leads to a least squares problem that can be solved to attain the series coefficients. Once these coefficients are known, the solution can be evaluated everywhere in the solution domain. To plot our theoretical predictions, $\Gamma$ is known a priori from (\ref{eq:circpred}) so that theoretical streamlines can be found in this manner.

However, for our aerofoil experiments, we wish to compare our experimentally attained circulation to the Kutta value of circulation. A modified solution scheme is necessary to deduce the Kutta circulation $\Gamma_K$, as it is unknown a priori and must be found as a part of the numerical solution. The Kutta value of circulation is chosen to enforce that the velocity does not diverge at the trailing edge of the body. 
The singular velocity can be avoided by enforcing the condition $dW(\zeta_p)/d\zeta_1=0$ through an appropriate choice of $\Gamma$, where $\zeta_p$ denotes the sharp corner pre-image. We now enforce this condition on the form given in (\ref{eq:Wform1}).  In particular, we consider the part of the potential that may become singular at the trailing edge,
\begin{equation}\label{eq:infpart}
    W_{\mathrm{inf}}(z)=\sum_{k=1}^{N_L}\frac{a^{(1)}_k}{\zeta_1(z)^k}+\mathrm{i}\frac{\Gamma_j}{2\pi}\log{\left(\zeta_1(z)\right)}.
\end{equation}
The normal component of the velocity in the flow will automatically be set to zero through the solution of the least squares problem that enforces (\ref{eq:Wbvp2}). Thus, in (\ref{eq:infpart}) we only need to set the tangential component of the velocity to be zero in order to regularize the velocity. Due to conformality, we can eliminate the tangential component of velocity directly in the $\zeta$-plane. 
Noting that $u=\overline{dW/dz}$, and the tangent vector on the unit circle is simply $t=i\zeta$, the vanishing tangential velocity condition becomes
\begin{equation}
    0=\mathrm{Re}\left\{\frac{dW_{\mathrm{inf}}}{d\zeta}\mathrm{i}\zeta\right\}\Bigg|_{\zeta=\zeta_p},
\end{equation}
from which we deduce the Kutta value of circulation,
\begin{equation}\label{eq:kutval}
    \Gamma_K=2\pi\sum_{k=1}^{N_L}k\mathrm{Im}\left\{\frac{a_k^{(1)}}{\zeta_p^k}\right\}.
\end{equation}
By substituting (\ref{eq:kutval}) in for $\Gamma$ in (\ref{eq:Wform1}), one finds a complex potential which enforces the Kutta condition. A modified least squares problem, compared to that when $\Gamma$ is specified independent of the coefficients $a_k^{(j)}$, can then be solved for the coefficients $a_k^{(j)}$ from which $\Gamma_K$ can be computed via (\ref{eq:kutval}). We have verified our numerical procedure by moving boundaries far away from the aerofoil and noting that the circulation converges to the value predicted by (\ref{eq:kutt}). 

\end{document}


\preprint{APS/123-QED}

\title{Supplemental Materials for\\
Characterizing dissipation in fluid-fluid displacement using constant-rate spontaneous imbibition}

\author{B. K. Primkulov}
\affiliation{Massachusetts Institute of Technology, Cambridge, USA}%

\author{J. Y. Y. Chui}
\affiliation{Massachusetts Institute of Technology, Cambridge, USA}%

\author{A. A. Pahlavan}
\affiliation{Princeton University, Olden St., Princeton, USA}%

\author{C. W. MacMinn}
\affiliation{University of Oxford, Oxford, United Kingdom}%

\author{R. Juanes}%
\email{juanes@mit.edu}
\affiliation{Massachusetts Institute of Technology, Cambridge, USA}%

\date{\today}

\maketitle

\section{Details of the experimental setup}
All of the experiments were conducted in Hilgenberg borosilicate glass tubes that are $75$\;mm in length and $290\;\mu$m in inner radius. The interfacial tensions of the oil--air and oil--water interfaces were $\gamma_{o}=22$\;mN/m and $\gamma_{ow}=13$\;mN/m, respectively. The dynamic contact angles of the water-oil interface in glass capillaries were measured under a microscope. The tubes were submerged into glycerol, which has a matching refractive index with the borosilicate glass in use (1.473). Contact angles were measured from the curvature of the interface, with parallax correction applied as in~\citep{Hoffman1975}.

Throughout this manuscript we assumed that Hagen--Poiseuille flow is maintained through the oil slug and that, therefore, the velocity profile is parabolic. This assumption was used to calculate the viscous drag and dissipation within the bulk of the oil slug. We confirmed the parabolic velocity profile within the oil slug through PIV tracing \citep{Thielicke2014PIVlabMATLAB}. In FIG.~\ref{fig:u_profile} we show that even for the shortest slug used in this study (2\;mm), the majority of the bulk space maintains the parabolic velocity profile.

\begin{figure}[!b]
 \centerline{\includegraphics[width=.4\linewidth]{Figure_s1.eps}}
 \caption{\label{fig:u_profile} PIV measurements of the velocity profile in spontaneously moving 2\;mm slug with 1000\;cSt viscosity. The plot is the 2D representation of a histogram, where color stands for the frequency. The data was collected over the entire length of the 2\;mm slug, over all frames. The figure demonstrates that even in the shortest slug used in this study (2\;mm), the majority of the bulk space maintains the parabolic velocity profile.}
\end{figure}

\section{Generalized Cox Equation}
In the main body of the manuscript we use the generalized Cox equation \cite{Cox1986TheFlow}
\begin{equation}
    g(\theta,M)-g(\theta_a,M) = \text{Ca}\,\Gamma,
\end{equation}
where $M$ is the ratio of the defending to invading fluid viscosities, $\Gamma=\ln(R/h_\text{micro})$ is the cut-off-length parameter near the contact line, and function $g(\theta,M)$ is
\begin{equation}
    g(\theta,M) = \int_{0}^{\theta} \frac{d\beta}{f(\beta,M)},
\end{equation}
and
\begin{equation}
    f(\beta,M) = \frac{2\sin\beta[M^2 (\beta^2-\sin^2\beta)+2M(\beta(\pi-\beta)+\sin^2\beta)+(\pi-\beta)^2-\sin^2\beta]} {M(\beta^2-\sin^2\beta)(\pi-\beta+\sin\beta\cos\beta)+((\pi-\beta)^2-\sin^2\beta)(\beta-\sin\beta\cos\beta)}.
\end{equation}

\section{Classical Imbibition}
FIG.~3 in the manuscript demonstrates that contact-line dissipation can be responsible for a significant portion of the energy loss in capillary-driven flow systems. To stress this point further, we return to the classical imbibition depicted in FIG.~1a of the manuscript. The need to account for contributions of the contact-line dynamics to the rate of classical imbibition has been the focus of a series of recent studies \citep{Bico2002Self-propellingSlugs, Delannoy2019TheRise, Hilpert2009EffectsSolutions, Hilpert2010ExplicitAngle, Heshmati2014ExperimentalSections}. We plot the evolution of the front position $z(t)$ for $50\;$cSt silicon oil in FIG.~\ref{fig:washburn_classical}. The classical Washburn scaling for $z(t)$ can be obtained by balancing $F_\text{bulk} = 8 \pi \mu_o z \dot{z}$ with $F_\text{cap}=2 \pi R \gamma_o \cos\theta_o = 2 \pi R \gamma_o (1- \frac{1}{2}(9\Gamma \frac{\mu_o \dot{z}}{\gamma_o})^{2/3})$ and neglecting the dynamic contact angle. Then the force balance reduces to
\begin{equation}
    \frac{4\mu_o}{R \gamma_o}z \dot{z} = 1.
    \label{eqn:washburn_classical}
\end{equation}
The solution to equation~\eqref{eqn:washburn_classical} is $z^2 = \frac{\gamma_o R}{2 \mu_o} t$, which differs from the early-time experimental data in FIG.~\ref{fig:washburn_classical}. A more complete description emerges by considering the dynamic contact angle
\begin{equation}
    \frac{4\mu_o}{R \gamma_o}z \dot{z} = 1 - \frac{1}{2}(9\Gamma \frac{\mu_o\dot{z}}{\gamma_o})^{2/3}.
    \label{eqn:washburn_corrected}
\end{equation}
Equation~\eqref{eqn:washburn_corrected} captures the dynamics of viscosity-dominated classical imbibition at both early- and late-times. At early times (when $z$ is small), $\Phi_\text{bulk}$ and $\Phi_\text{cl}$ are comparable (see FIG.~\ref{fig:washburn_classical}) and therefore the dynamics is best described by including both dissipation sources. At late times, the liquid front slows and $\theta_{o}$ approaches $\theta_{o,a}$, making $\Phi_\text{cl}$ negligible. As a result, the experimental $z(t)$ approaches the $z \sim t^{1/2}$ scaling (FIG.~\ref{fig:washburn_classical}).

\begin{figure}
 \centerline{\includegraphics[width=0.4\linewidth]{Figure_s2.eps}}
 \caption{\label{fig:washburn_classical} Evolution of $z(t)$ during classical imbibition of 50\;cSt silicon oil depicted in FIG.~1a of the manuscript. Here the black line represents the classical Washburn solution [Eq.~\eqref{eqn:washburn_classical}], the red line represents the solution corrected for dynamic contact angle [Eq.~\eqref{eqn:washburn_corrected}]. The ratio of contact-line to total dissipation is denoted with a colormap.}
\end{figure}

\bibliography{references.bib}